\documentclass{caosp}
\usepackage{graphicx}
\articleNo{230}
\pubyear{2013}
\volume{43}
\volnumber{2}
\firstpage{94}
\received{June 19, 2013}
\accepted{October 16, 2013}
\begin{document}
\title{Optical observations of SN 2011fe}
\author{
        D.Yu.\,Tsvetkov\inst{1}
      \and
        S.Yu.\,Shugarov\inst{1,2}
      \and
        I.M.\,Volkov\inst{1,2}
      \and
        V.P.\,Goranskij\inst{1}
      \and
        N.N.\,Pavlyuk\inst{1}
      \and
        N.A.\,Katysheva\inst{1}
      \and
       E.A.\,Barsukova\inst{3}
      \and
      A.F.\,Valeev\inst{3}
       }
\institute{
           Sternberg Astronomical Institute, M.V.\,Lomonosov Moscow State
          University,
          Universitetskii pr. 13, 119992 Moscow, Russia\\
           \email{tsvetkov@sai.msu.su}
         \and
           \lomnica
         \and
    Special Astrophysical Observatory of the Russian Academy of
     Sciences, Nizhniy Arkhyz, Karachai-Cherkesia, 369167 Russia
         }

\hauthor{D.Yu.\,Tsvetkov {\it et al.}}
\date{June 19, 2013}
\maketitle
\begin{abstract}
We present {\it UBVRI} photometry of the supernova 2011fe in M101,
obtained in the interval of 1 -- 652 days after its discovery, as well as one
spectrum, taken 105 days after the brightness maximum in the $B$ band.
We derived  parameters of the light curves,
constructed the colour curves and a "quasi--bolometric" light curve.
The light curves, colour evolution and spectrum of the object indicate that
SN 2011fe belongs to the "normal" subset of
type Ia supernovae. It is practically identical to a well-studied
"normal" SN Ia 2003du.

\end{abstract}
\section{Introduction}

Supernova (SN) 2011fe, located at $\alpha=14^{\rm h}03^{\rm m}05^{\rm s}.81,
\delta=+54^{\circ}16'25''.4$ (2000.0) in the galaxy M101 (NGC 5457), 
was discovered
by the Palomar Transient Factory
on UT 2011 Aug 24.2, less than one day after its explosion
(Nugent {\it et al.}, 2011).
It is the closest and brightest
type Ia SN since SN 1972E, so it provides an unprecedented opportunity for
numerous follow-up studies.

Li {\it et al}. (2011) checked the archival HST images of the site and
found out that a luminous red giant cannot be the companion to the
SN progenitor. This conclusion was confirmed by
early X-ray and radio observations
(Horesh {\it et al.}, 2012; Chomiuk {\it et al.}, 2012).
The results of photometric monitoring were presented by
Vinko {\it et al.} (2012,
hereafter V12),
Richmond and Smith (2012, hereafter RS12) and Munari {\it et al.} (2013,
hereafter M13). Pereira {\it et al.} (2013, hereafter P13)
reported spectrophotomeric observations and derived synthetic
{\it UBVRI} magnitudes based on these data.
Infrared photometry was presented by Matheson {\it et al.} (2012).
Optical spectral evolution was investigated by Parrent {\it et al.} (2012),
Smith {\it et al.} (2011), Patat {\it et al.} (2013) and
Shappee {\it et al.} (2013).
This bright nearby event should provide a wealth of
information on the nature of thermonuclear supernovae.

\section{Observations}

We present here photometry of SN 2011fe in the
{\it UBVRI} passbands obtained at five sites,
starting one day after the discovery and
continuing for a period of 652 days.
Most of the data were obtained
at the Star\'a Lesn\'a Observatory
of the Astronomical Institute of the Slovak Academy of Sciences.
The other observing sites were the Crimean Laboratory of the Sternberg
Astronomical Institute (SAI)(Nauchniy, Crimea, Ukraine); the
Simeiz Observatory of the Crimean Astrophysical Observatory (Simeiz, Crimea,
Ukraine); the Moscow Observatory of SAI (Moscow, Russia); the
Special Astrophysical Observatory of RAS (Nizhniy Arkhyz, Russia).
A list of the observing facilities is given
in Table\,\ref{t1}.

\begin{table}
\begin{center}
\caption{Telescopes and detectors employed for the observations.}
\label{t1}
\begin{tabular}{ccccccc}
\hline\hline
Tele-  & Location &
Aperture &  CCD  & Filters & Scale & FoV  \\
scope &  & [m] & camera  &  &   [arcsec & [arcmin] \\
 code  &  &     &  &  &     pixel$^{-1}$]             &     \\
\hline
T50  & Star\'a             & 0.5 & SBIG       & $UBVR_CI_C$  &1.12 & 20.4x13.7\\
     & Lesn\'a,            &     & ST-10   &          &      &    \\
     & Slovakia            &     & XME           &          &      &   \\
T15  &   -"-               & 0.15&   -"-     & $BVR_CI_C$   & 1.26 &22.9x15.5\\
T60  &   -"-               & 0.6 & VersArray & $UBVR_CI_J$& 1.38 & 5.9  \\
     &                     &    &   F512     &          &      &      \\
C60  & Nauchniy,          & 0.6 & Apogee     & $UBVR_CI_J$& 0.71 & 6.1 \\
     & Crimea,            &      & AP-47p    &         &      &     \\
     &  Ukraine            &     &           &          &      &    \\
S100 &  Simeiz,           & 1.0 & VersArray  &$UBVR_CI_J$ & 0.65 & 7.2 \\
     &  Crimea,            &    &  B1300     &          &     &     \\
     & Ukraine             &     &           &          &      &      \\ 
M70  & Moscow,            & 0.7 & Apogee     & $UBVR_CI_J$& 0.64 & 5.5 \\
     & Russia              &    &  AP-7p     &          &      &     \\
N100 & Nizhniy            & 1.0 & EEV        & $UBVR_CI_C$  & 0.48 & 8.3 \\
     & Arkhyz,            &     &  42-40     &          &      &      \\
     &  Russia             &     &           &          &      &      \\
N600 &  -"-                & 6.0 &  -"-      &   $V$    & 0.36 & 6.3  \\
\hline\hline         
\end{tabular}        
\end{center}         
\end{table} 

\begin{figure}
\centerline{\includegraphics[width=11cm,clip=]{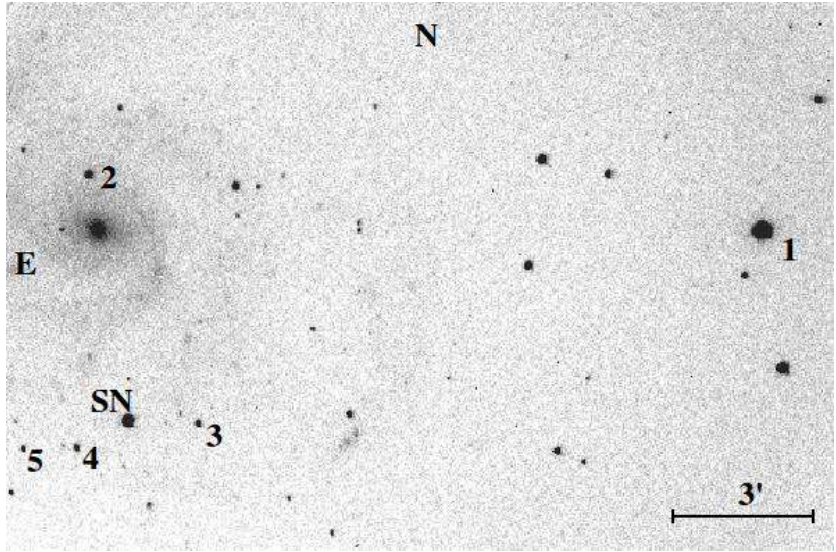}}
\caption{The image of SN2011fe and local standard stars, obtained by
the T50 telescope in the $R$ band.}
\label{f1}
\end{figure}

The standard image reductions and photometry were made using the
IRAF\footnotemark .
\footnotetext{IRAF is distributed by the National Optical Astronomy Observatory,
which is operated by AURA under cooperative agreement with the
National Science Foundation.}
The magnitudes of the SN
were derived by an aperture photometry or a PSF-fitting relatively to
a sequence of local standard stars.
The CCD image of SN 2011fe and local standard stars is presented
in Fig.\,\ref{f1}.

The {\it BVRI} magnitudes of stars No.2-5 were taken from Henden
{\it et al.}
(2012). Star No.1 (HD 122601) was calibrated on 5 photometric nights
using the telescopes
C60 and M70. The resulting magnitudes are $U=9.44\pm0.03; B=9.44\pm0.02;
V=9.02\pm0.01; R=8.76\pm0.01;$ and $I=8.52\pm0.02$.
The $U$-band magnitudes of the stars No 2, 3 and 4, measured relatively
to the star No.1,
are 13.44$\pm$0.04, 14.68$\pm$0.05 and 15.02$\pm$0.05, respectively.

The photometry of the SN, obtained by different telescopes, is given in
Table\,\ref{t2}, Table\,\ref{t3}, and Table\,\ref{t4}.

\begin{table}
\begin{center}
\tabcolsep=5pt
\caption{$UBVRI$ magnitudes of SN2011fe from the 50-cm reflector
at Star\'a Lesn\'a.}
\label{t2}
\begin{tabular}{ccccccccccc}
\hline\hline
JD$-$ & $U$ & $\sigma_U$ & $B$ & $\sigma_B$ & $V$ & $\sigma_V$ &
$R$ & $\sigma_R$ & $I$ & $\sigma_I$ \\
2450000 & & & & & & & & & & \\
\hline
5799.38& 14.08 & 0.07& 14.36 & 0.03 & 13.98 & 0.04 & 13.91 & 0.05 & 13.94 & 0.04 \\
5800.29& 13.17 & 0.06& 13.51 & 0.05 & 13.24 & 0.03 & 13.18 & 0.03 & 13.22 & 0.03 \\ 
5801.29& 12.51 & 0.06& 12.83 & 0.04 & 12.66 & 0.04 & 12.62 & 0.04 & 12.55 & 0.03 \\ 
5803.27& 11.46 & 0.08& 11.83 & 0.03 & 11.78 & 0.02 & 11.69 & 0.03 & 11.65 & 0.03 \\ 
5806.29& 10.27 & 0.27& 10.79 & 0.03 & 10.90 & 0.02 & 10.77 & 0.03 & 10.76 & 0.02 \\ 
5806.30& 10.28 & 0.03& 10.77 & 0.04 & 10.82 & 0.02 & 10.72 & 0.02 & 10.75 & 0.04 \\ 
5807.28& 10.07 & 0.03& 10.58 & 0.02 & 10.65 & 0.01 & 10.52 & 0.02 & 10.59 & 0.03 \\ 
5808.27&  9.91 & 0.03& 10.45 & 0.06 & 10.49 & 0.02 & 10.37 & 0.02 & 10.44 & 0.02 \\ 
5809.31&  9.79 & 0.03& 10.31 & 0.05 & 10.37 & 0.02 & 10.29 & 0.03 & 10.35 & 0.02 \\ 
5815.36&  9.55 & 0.03&  9.97 & 0.06 &  9.99 & 0.03 & 10.00 & 0.03 & 10.34 & 0.05 \\ 
5817.63&  9.70 & 0.04& 10.03 & 0.03 & 10.05 & 0.01 & 10.08 & 0.04 & 10.37 & 0.03 \\ 
5818.24&  9.67 & 0.03& 10.00 & 0.04 & 10.02 & 0.02 &  9.99 & 0.04 & 10.44 & 0.04 \\ 
5820.24&  9.82 & 0.03& 10.14 & 0.02 & 10.08 & 0.02 & 10.09 & 0.03 & 10.57 & 0.03 \\ 
5830.24& 10.94 & 0.03& 11.08 & 0.02 & 10.68 & 0.01 & 10.73 & 0.01 & 10.97 & 0.02 \\ 
5833.23& 11.39 & 0.03& 11.42 & 0.02 & 10.79 & 0.01 & 10.75 & 0.01 & 10.85 & 0.02 \\ 
5836.29& 11.76 & 0.03& 11.76 & 0.02 & 10.91 & 0.01 & 10.73 & 0.01 & 10.75 & 0.02 \\ 
5839.65& 12.22 & 0.07&       &      & 11.13 & 0.02 & 10.85 & 0.02 & 10.67 & 0.03 \\ 
5844.26& 12.50 & 0.06& 12.56 & 0.04 & 11.41 & 0.03 & 11.06 & 0.02 & 10.73 & 0.03 \\ 
5856.22& 13.09 & 0.04& 13.13 & 0.03 & 12.05 & 0.03 & 11.78 & 0.02 & 11.50 & 0.03 \\ 
5861.26& 13.25 & 0.05& 13.26 & 0.03 & 12.28 & 0.02 & 12.00 & 0.02 & 11.77 & 0.04 \\ 
5862.68& 13.16 & 0.06& 13.23 & 0.06 & 12.25 & 0.05 & 12.01 & 0.04 & 11.82 & 0.04 \\ 
5863.68& 13.16 & 0.06& 13.22 & 0.04 & 12.14 & 0.03 &       &      &       &      \\
5867.68& 13.21 & 0.12& 13.23 & 0.03 & 12.28 & 0.04 & 12.15 & 0.03 & 11.99 & 0.05 \\ 
5868.64& 13.30 & 0.06& 13.32 & 0.02 & 12.38 & 0.02 & 12.19 & 0.01 & 12.06 & 0.02 \\ 
5871.69& 13.28 & 0.06& 13.36 & 0.04 & 12.50 & 0.02 & 12.33 & 0.03 & 12.22 & 0.03 \\ 
5872.17& 13.35 & 0.06& 13.38 & 0.02 & 12.50 & 0.02 & 12.32 & 0.02 & 12.28 & 0.02 \\ 
5874.26& 13.56 & 0.06& 13.50 & 0.04 & 12.65 & 0.05 & 12.50 & 0.08 & 12.40 & 0.06 \\ 
5876.26& 13.29 & 0.08& 13.58 & 0.02 & 12.70 & 0.03 & 12.51 & 0.04 & 12.50 & 0.05 \\ 
5878.70& 13.57 & 0.07& 13.57 & 0.04 & 12.74 & 0.04 & 12.58 & 0.03 & 12.58 & 0.04 \\ 
5883.67& 13.73 & 0.06& 13.64 & 0.05 & 12.86 & 0.03 & 12.71 & 0.03 & 12.76 & 0.02 \\ 
5889.70& 13.69 & 0.09& 13.70 & 0.02 & 13.03 & 0.05 & 12.90 & 0.02 & 12.93 & 0.06 \\ 
5893.71& 14.08 & 0.30& 13.72 & 0.06 & 13.09 & 0.03 & 13.06 & 0.05 & 13.28 & 0.09 \\ 
5925.63& 14.75 & 0.10& 14.19 & 0.02 & 13.91 & 0.02 & 14.04 & 0.01 & 14.22 & 0.02 \\ 
5933.72& 15.06 & 0.16& 14.23 & 0.06 & 14.00 & 0.03 & 14.27 & 0.04 & 14.49 & 0.07 \\ 
5956.66& 15.69 & 0.08& 14.66 & 0.04 & 14.51 & 0.03 & 14.88 & 0.03 & 14.96 & 0.02 \\ 
5967.45& 16.24 & 0.15& 14.79 & 0.03 & 14.74 & 0.03 & 15.12 & 0.02 & 15.15 & 0.04 \\ 
5974.41& 15.70 & 0.14& 14.89 & 0.03 & 14.86 & 0.02 & 15.32 & 0.04 & 15.16 & 0.04 \\ 
5981.41& 16.19 & 0.19& 15.00 & 0.03 & 14.97 & 0.02 & 15.48 & 0.03 & 15.32 & 0.03 \\ 
5987.61& 16.45 & 0.15& 15.10 & 0.02 & 15.10 & 0.03 & 15.64 & 0.04 & 15.43 & 0.03 \\ 
6019.60&       &     & 15.54 & 0.03 & 15.61 & 0.02 & 16.26 & 0.06 & 15.80 & 0.05 \\ 
6028.53&       &     & 15.70 & 0.10 & 15.71 & 0.09 & 16.07 & 0.29 &       &      \\
6045.59& 17.33 & 0.18& 15.99 & 0.05 & 16.01 & 0.03 & 16.83 & 0.07 & 16.43 & 0.19 \\ 
\hline\hline         
\end{tabular}        
\end{center}        
\end{table}
\begin{table} 
\begin{center}
\tabcolsep=5pt
\caption{Photometry of SN2011fe from the 60-cm reflector 
at Crimea.}
\label{t3}
\begin{tabular}{ccccccccccc}
\hline\hline
JD$-$ & $U$ & $\sigma_U$ & $B$ & $\sigma_B$ & $V$ & $\sigma_V$ &
$R$ & $\sigma_R$ & $I$ & $\sigma_I$ \\
2450000 &  & & & & & & & & &  \\
\hline
5799.35&       &      &  14.33 & 0.03 & 14.03 & 0.02&  13.91 & 0.02 & 13.72 & 0.03\\  
5800.31& 13.54 & 0.10 &  13.50 & 0.04 & 13.23 & 0.02&  13.16 & 0.01 & 12.97 & 0.03\\  
5801.30&       &      &  12.83 & 0.02 & 12.68 & 0.01&  12.57 & 0.01 & 12.42 & 0.03\\ 
5802.31&       &      &  12.32 & 0.02 & 12.16 & 0.02&  12.11 & 0.01 & 11.95 & 0.03\\ 
5803.31&       &      &  11.81 & 0.01 & 11.75 & 0.01&  11.64 & 0.01 & 11.54 & 0.02\\ 
5804.32&       &      &  11.45 & 0.04 & 11.42 & 0.03&  11.27 & 0.01 & 11.20 & 0.03\\ 
5807.27&       &      &  10.64 & 0.08 & 10.69 & 0.02&  10.56 & 0.03 & 10.57 & 0.06\\ 
5807.33& 10.14 & 0.08 &  10.65 & 0.02 & 10.66 & 0.02&  10.55 & 0.01 & 10.57 & 0.02\\  
5808.36&       &      &  10.49 & 0.01 & 10.51 & 0.02&  10.39 & 0.02 & 10.42 & 0.02\\ 
5811.31&  9.48 & 0.05 &  10.07 & 0.03 & 10.13 & 0.01&  10.11 & 0.01 & 10.21 & 0.02\\  
5879.16& 13.86 & 0.09 &  13.58 & 0.02 & 12.72 & 0.04&  12.60 & 0.03 & 12.41 & 0.05\\  
5916.59&       &      &        &      & 13.65 & 0.03&  13.70 & 0.02 & 13.69 & 0.03\\ 
6153.30&       &      &  17.92 & 0.08 & 17.76 & 0.04&  18.30 & 0.04 &       &    \\
6155.26&       &      &  17.69 & 0.05 & 17.79 & 0.03&  18.28 & 0.04 &       &    \\
6156.30&       &      &        &      & 17.87 & 0.06&  18.34 & 0.10 &       &    \\
6160.31&       &      &        &      & 17.85 & 0.03&  18.36 & 0.03 &       &    \\
6166.28&       &      &        &      & 17.96 & 0.03&  18.42 & 0.04 &       &    \\
6172.25&       &      &  18.04 & 0.12 & 18.00 & 0.03&  18.52 & 0.05 &       &    \\
6173.25&       &      &        &      & 18.05 & 0.03&  18.47 & 0.04 &       &    \\
6174.27&       &      &        &      & 18.09 & 0.02&  18.59 & 0.04 &       &    \\
6176.26&       &      &        &      & 18.18 & 0.03&  18.65 & 0.07 &       &    \\
6177.28&       &      &        &      & 18.10 & 0.03&  18.67 & 0.05 &       &    \\
6249.62&       &      &        &      & 19.32 & 0.04&  19.73 & 0.07 &       &    \\
6250.61&       &      &  19.38 & 0.06 & 19.38 & 0.06&  19.62 & 0.07 &       &    \\
6252.65&       &      &        &      & 19.40 & 0.10&        &      &       &    \\  
\hline\hline         
\end{tabular}        
\end{center}         
\end{table}  
\begin{table} 
\begin{center}
\tabcolsep=5pt
\caption{Photometry of SN2011fe at 6 telescopes.}
\label{t4}
\begin{tabular}{cccccccccccl}
\hline\hline
JD$-$ & $U$ & $\sigma_U$ & $B$ & $\sigma_B$ & $V$ & $\sigma_V$ &
$R$ & $\sigma_R$ & $I$ & $\sigma_I$ & Tel. \\
2450000 & & & & & & & & & & \\
\hline
5808.30 &       &    &       &       & 10.50 & 0.01 & 10.38 & 0.02 & 10.42 & 0.02& T15 \\
5817.25 &  9.80 & 0.04& 10.05 & 0.01 & 10.04 & 0.01 & 10.04 & 0.01 &       &    & S100 \\
5818.24 &  9.92 & 0.04& 10.09 & 0.01 & 10.05 & 0.01 &       &     &       &    & S100 \\
5819.24 &  9.92 & 0.04& 10.15 & 0.01 & 10.07 & 0.01 & 10.06 & 0.01 &       &    & S100\\
5820.24 & 10.10 & 0.04& 10.20 & 0.01 & 10.10 & 0.01 &       &     &       &    & S100\\
5821.23 & 10.15 & 0.04& 10.24 & 0.01 & 10.12 & 0.01 &       &     &       &    & S100\\
5822.20 & 10.31 & 0.04& 10.32 & 0.01 & 10.18 & 0.01 & 10.23 & 0.01 &       &    & S100\\
5823.25 & 10.33 & 0.04& 10.40 & 0.01 & 10.20 & 0.01 & 10.30 & 0.01 &       &    & S100\\
5825.22 & 10.56 & 0.04& 10.55 & 0.01 & 10.32 & 0.01 & 10.43 & 0.01 &       &    & S100\\
5825.22 & 10.42 & 0.05& 10.65 & 0.01 & 10.30 & 0.03 & 10.44 & 0.03 & 10.89 & 0.01& N100\\
5830.22 & 11.27 & 0.04& 11.11 & 0.01 & 10.63 & 0.01 & 10.73 & 0.01 &       &    & S100\\
5831.22 & 11.39 & 0.08& 11.22 & 0.01 & 10.72 & 0.01 & 10.75 & 0.01 &       &    & S100\\
5832.21 & 11.70 & 0.05& 11.36 & 0.01 & 10.76 & 0.02 & 10.75 & 0.02 &       &    & S100\\
5835.22 &       &    & 11.61 &  0.03 & 10.86 & 0.02 & 10.75 & 0.04 & 10.77 & 0.03& T15\\ 
5852.25 &       &    &       &      & 11.91 & 0.02 & 11.59 & 0.02 & 11.27 & 0.02& T15\\ 
5852.26 & 13.20 &0.08& 13.07 & 0.01 & 11.90 & 0.02 & 11.63 & 0.02 & 11.18 & 0.03& T60\\
5853.22 &       &    & 13.06 & 0.05 &       &     &       &     &       &    & T60\\
5853.22 &       &    &       &     & 11.94 & 0.02 & 11.66 & 0.02 & 11.33 & 0.02& T15\\ 
5856.23 &       &    &       &     & 12.06 & 0.02 & 11.79 & 0.03 & 11.51 & 0.03& T15\\
5856.23 &       &    & 13.16 & 0.02 &       &     &       &     &       &    & T60\\
5919.61 &       &    &       &     & 13.64 & 0.04 &       &     &       &    & N600\\
5953.49 &       &    & 14.61 & 0.03 & 14.48 & 0.02 & 14.65 & 0.02 & 14.54 & 0.03& M70\\
5954.69 &       &    & 14.63 & 0.05 & 14.59 & 0.03 & 14.81 & 0.02 & 14.63 & 0.04& T60\\
5955.60 & 15.71 &0.08& 14.58 & 0.02 & 14.58 & 0.02 & 14.80 & 0.02 & 14.56 & 0.03& T60\\
5987.61 &       &    & 15.17 & 0.04 & 15.19 & 0.02 & 15.61 & 0.03 & 15.59 & 0.04& T15\\
6014.35 &       &    & 15.50 & 0.03 & 15.46 & 0.03 & 15.97 & 0.04 & 15.60 & 0.05& M70\\
6022.45 & 17.17 &0.11& 15.65 & 0.02 & 15.77 & 0.03 & 16.21 & 0.02 & 15.73 & 0.03& S100\\
6025.40 &       &    & 15.69 & 0.03 & 15.84 & 0.06 & 16.31 & 0.06 &       &   & S100\\
6194.23 &       &    & 18.16 & 0.05 & 18.39 & 0.04 & 18.86 & 0.05 &       &    & S100\\
6196.20 &       &    &       &     & 18.45 & 0.06 & 18.81 & 0.09 &       &    & S100\\
6199.20 &       &    & 18.30 & 0.08 & 18.46 & 0.07 & 18.88 & 0.08 &       &    & S100\\
6387.48 &       &    & 21.20 & 0.14 &       &       &      &      &       &    & S100\\
6441.31 &       &    &       &      & 21.23 & 0.07  & 22.00 & 0.12 &       &    & S100\\
6442.34 &       &    &       &      &       &       & 22.07 & 0.11 &       &    & S100\\ 
6446.28 &       &    &       &      & 21.15 & 0.17  &       &      &       &    & S100\\        
6450.36 &       &    &       &      & 21.36 & 0.08  &       &      &       &    & S100\\               
\hline\hline         
\end{tabular}        
\end{center}         
\end{table}  

The photometry was transformed to the standard Johnson-Cousins
system by means of instrument colour-terms, determined from
observations of standard star clusters. The procedure was described
in details by Elmhamdi {\it et al.} (2011), Tsvetkov
{\it et al.} (2008), and Tsvetkov {\it et al.} (2006).
The type of $R$- and $I$-filters is indicated in Table\,\ref{t1}.
We transformed the photometry in the $R, I$-bands to Cousins
system, so $R$ and $I$ are equivalent to $R_C$, $I_C$.
The reported errors were computed by adding in quadratures the
fitting errors returned by IRAF tasks and the uncertainties of
local standards calibration.
The brightness of the SN near the maximum presented a significant
difficulty for photometry, as no sufficiently bright comparison
stars could be found close to the object. The images at the
telescopes with large FoV (T50, T15) were obtained with the SN near
the eastern edge, and star No 1 near the western edge. Such a position
of the FoV allowed estimation of SN brightness on images with short exposures,
but could result in field errors, which were difficult to account for.
While observing at telescopes with smaller
FoV, we sometimes used the star No 2 as the main standard, or fainter but
nearer stars No 3, 4, 5. At late stages only these fainter stars were used.
The background of the host galaxy was negligible
for most of the monitoring period, this is evident from our images and
from the analysis presented by RS12. However, for the images
obtained later than 570 days after the $B$-maximum, we applied
image subtraction using SDSS\footnotemark\  images of the field around
\footnotetext{http://www.sdss.org}
the SN position.

The spectroscopic observations were carried out at the 6-m BTA telescope
of SAO RAS (N600) on UT 2011 December 24.13. The focal reducer SCORPIO
with the grism VPHG1200G provided the wavelength range of 4044--5858 \AA\,  
with a
dispersion of 0.88 \AA pixel$^{-1}$. The spectra were bias and
flat-field corrected, extracted and wavelength calibrated in ESO/MIDAS.
The spectrophotometric standard AGK+81$^{\circ}$266 was used for flux calibrated
spectra, but the night was not photometric and the absolute
flux values may have significant errors.

\section{Light and colour curves}

\begin{figure}
\centerline{\includegraphics[width=11cm,clip=]{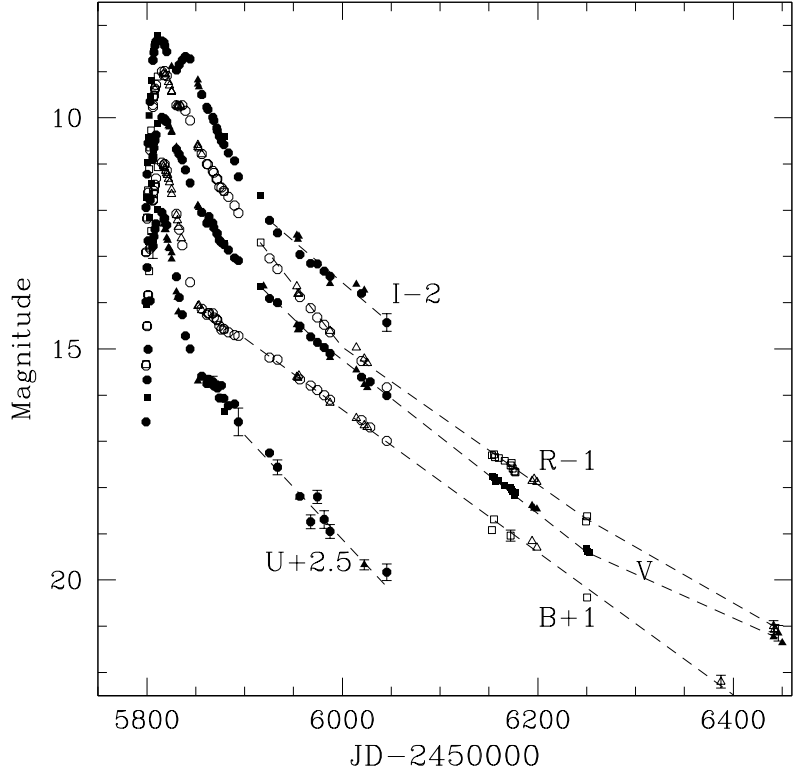}}
\caption{The light curves of SN 2011fe in the {\sl UBVRI} bands.
Circles show the data from T50, squares -- from C60, triangles --
from other telescopes. For clarity of presentation, the
data in {\sl BR} bands are plotted with open symbols and the
curves are shifted in magnitude. The shifts
for every band are reported.
The error bars are plotted only when they exceed the size of
a symbol.
Dashed lines present the linear fits to the late phases.}
\label{f2}
\end{figure}

The light curves of SN 2011fe are presented in Fig.\,\ref{f2}.
The results for all the telescopes are in a
fairly good agreement, the largest
differences are found in the $U$-band. The shape of the light
curves is typical for SNe Ia.

We fitted the light curves with cubic splines and determined the
dates and magnitudes of maximum light in different bands and the
decline rate parameters $\Delta m_{15}$. These
are reported in Table\,\ref{t5}.
The errors of magnitudes at maximum and $\Delta m_{15}$ are about
0.02--0.05 mag, the errors of their times are 0.1--0.2 days.
The results are consistent with the data reported by other authors (V12, RS12,
M13).
At late phases we used linear fits
for the light curves. The resulting parameters are listed in Table\,\ref{t6}.
The rates of decline in the interval JD 2456000--6250 (which corresponds
to phases 180--430 days after the $B$-maximum) in the {\sl BVR-} bands
are significantly greater than the mean values for SNe Ia (Lair {\it et al.},
2006). The comparison with the data for SN 2003du, which has similar
$\Delta m_{15}$ (Stanishev {\it et al.}, 2007), reveals that the
rate of decline for SN 2011fe is larger in the $B$- and $V$-band and
lower in the $R$-band.
After JD 2456250 (phase 430 days) the decline
in the {\sl VR}- bands slows down. Similar behaviour of the $V$-band light
curve can
be noticed for SN 2000E (Lair {\it et al.}, 2006) and SN 1992A
(Cappellaro {\it et al.}, 1997).

\begin{table}
\begin{center}
\caption{Dates and magnitudes of maximum light and the
decline rate parameters in different passbands.}
\label{t5}
\begin{tabular}{ccrc}
\hline\hline
Band & JD$-$2455000   & mag & $\Delta m_{15}$ \\
\hline
\multispan{2} Primary maximum & \\
$U$  & 813.3 &  9.53 &  1.36 \\
$B$  & 815.1 & 10.01 &  1.10 \\
$V$  & 815.9 & 10.02 &  0.67 \\
$R$  & 815.4 &  9.97 &  0.75 \\
$I$  & 812.8 & 10.21 &  0.76 \\
\multispan{2} Secondary maximum & \\
$I$  & 839.9 & 10.68 &       \\
\hline\hline
\end{tabular}
\end{center} 
\end{table}  
\begin{table}
\begin{center}
\caption{Rates of brightness decline (in mag/100$^d$) at different late stages
of SN 2011fe evolution.}
\label{t6}
\begin{tabular}{cccc}
\hline\hline
Band & JD 2455900-6000 & JD 2456000-6250 & JD 2456250-6450 \\
\hline
$U$  & 2.26 $\pm$ 0.12 &  &   \\
$B$  & 1.51 $\pm$ 0.02 & 1.64 $\pm$ 0.07  & \\
$V$  & 2.05 $\pm$ 0.03 & 1.60 $\pm$ 0.01 & 0.99 $\pm$ 0.03 \\
$R$  & 2.62 $\pm$ 0.03 & 1.54 $\pm$ 0.01 & 1.23 $\pm$ 0.05 \\
$I$  & 1.85 $\pm$ 0.03 &  &  \\
\hline\hline
\end{tabular}
\end{center} 
\end{table}  

\begin{figure}
\centerline{\includegraphics[width=11cm,clip=]{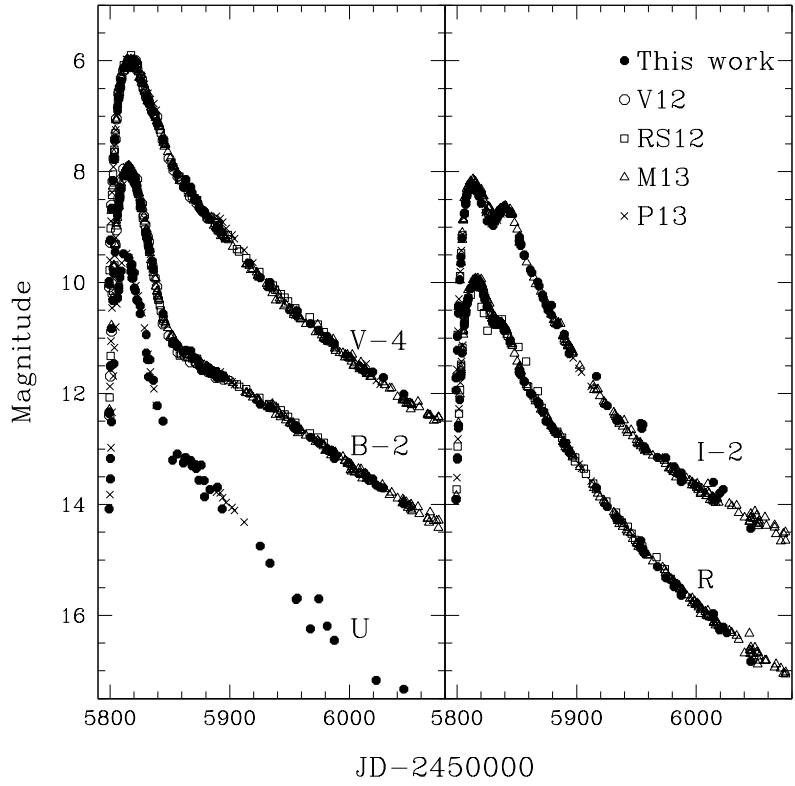}}
\caption{The photometric data for SN 2011fe from different authors.}
\label{f3}
\end{figure}

Fig.\,\ref{f3} presents the comparison of photometry by different
authors near the maximum brightness. The superposed light
curves are in a good agreement.
We may only note some outlying points by RS12 in the $R$-band and
for our data in the $I$-band.
We computed mean differences between our data and other four
main sets (RS12, M13, V12, P13)
in different bands. We found the best consistency with
the RS12 and M13 data, where the mean difference does not
exceed 0.04 mag. The agreement with the V12
and P13 data is worse. The maximum mean difference
is 0.13 mag.

\begin{figure}
\centerline{\includegraphics[width=11cm,clip=]{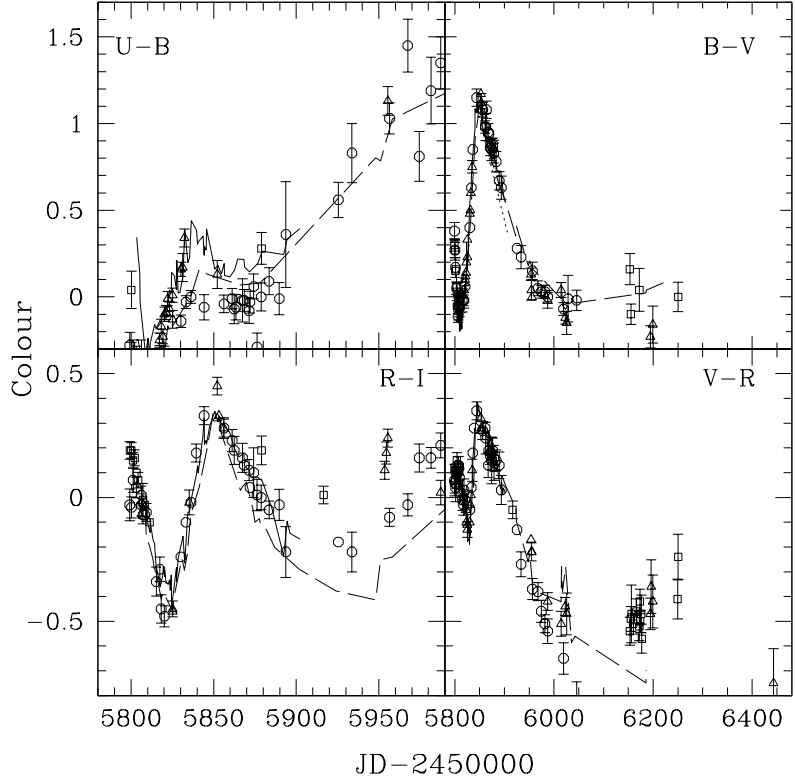}}
\caption{The colour curves of SN 2011fe. The meaning of the symbols
is the same as in Fig. 2. The colour curves of SN 2003du (dashed
line) and SN 2005cf (solid line) are plotted for comparison.
The curves for SN 2005cf were corrected for $E(B-V)=0.097$.
The dotted line on the $B-V$ diagram is the Lira-Phillips relation.}
\label{f4}
\end{figure}

As seen from colour curves, presented in Fig.\,\ref{f4},
the data from different telescopes are in a fairly good agreement.
Some inconsistencies and large errors are evident for the $U-B$ colour
and for the late-time $R-I$ data.
The colour evolution is typical for SN Ia, this is confirmed by
comparison with the colour curves for "normal", unreddened SNe Ia
with nearly
the same value of $\Delta m_{15}$: SN 2003du (Stanishev {\it et al.}, 2007)
and SN 2005cf (Pastorello {\it et al.}, 2007).
The $B-V$ colour curve is also compared with the "Lira-Phillips relation"
(Phillips {\it et al.}, 1999), showing the time dependence of $B-V$
in the phase interval 30-90 days for most of SNe Ia that suffered no
extinction.
It is obvions, that the interstellar extinction towards SN 2011fe
is very low.

\begin{figure}
\centerline{\includegraphics[width=11cm,clip=]{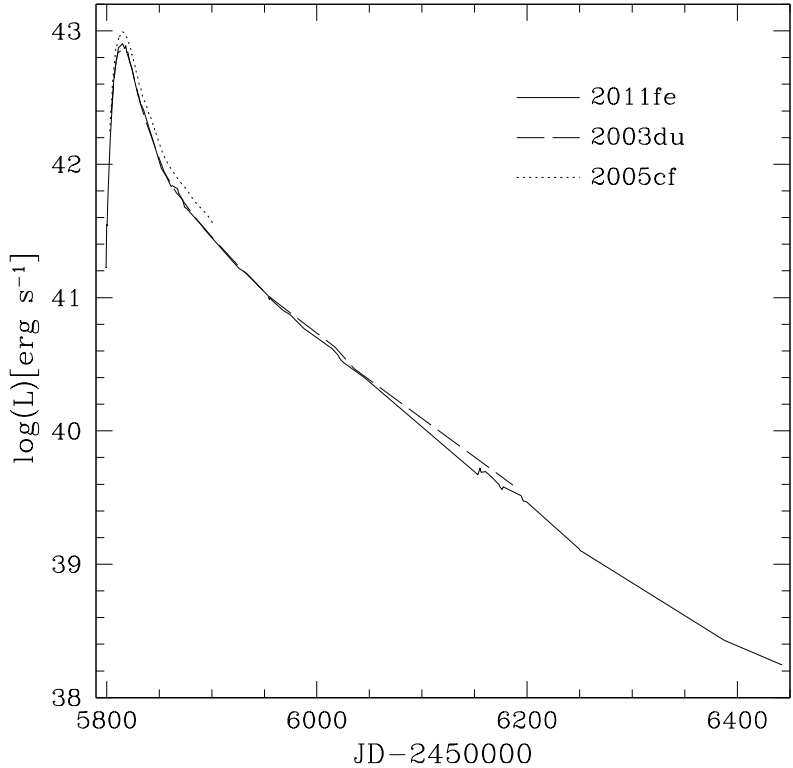}}
\caption{The "quasi-bolometric" light curve of SN 2011fe,
compared to the light curves of SNe 2003du and 2005c.}
\label{f5}
\end{figure}

The "quasi-bolometric" light curve for SN 2011fe, constructed
by integrating the flux from $U$- to $I$-bands, assuming distance
modulus $\mu=29.04$ (Shappee and Stanek, 2011) and extinction $E(B-V)=0.009$
(Schlegel {\it et al.}, 1998) is shown in Fig.\,\ref{f5}.
On the dates when observations in some
bands were missing, we interpolated or extrapolated the colour curves
to estimate the colour of SN and then calculated the missing magnitudes.
We
compare SN 2011fe to SNe 2003du and 2005cf, for which the
"quasi-bolometric" light curves were
constructed
analogously. The adopted distance modulus and extinction for these
objects are:
$\mu=32.50$, $E(B-V)=0.01$ for SN 2003du (Tsvetkov {\it et al.}, 2011) and
$\mu=32.51$, $E(B-V)=0.097$ for SN 2005cf (Pastorello {\it et al.}, 2007).
The "quasi-bolometric" light curves
for SN 2011fe and 2003du are nearly identical, while SN 2005cf is slightly
brighter. The rate of decline for bolometric luminosity
is 1.72$\pm$0.03 mag/100$^d$
in the
interval JD 2455900-2456100,
1.55$\pm$0.04 in the
interval JD 2456150-2456250, and
1.33$\pm$0.04
in the interval
JD 2456250-2456450. The gradual reduction of the
rate of brightness decline is evident from this data.
Intergating the "quasi-bolometric" light curve of SN 2011fe over time we
find that the total optical energetic output of the SN 
equals $2.05\times10^{49}$ ergs. 

\section{Spectrum}

\begin{figure}
\centerline{\includegraphics[width=11cm,clip=]{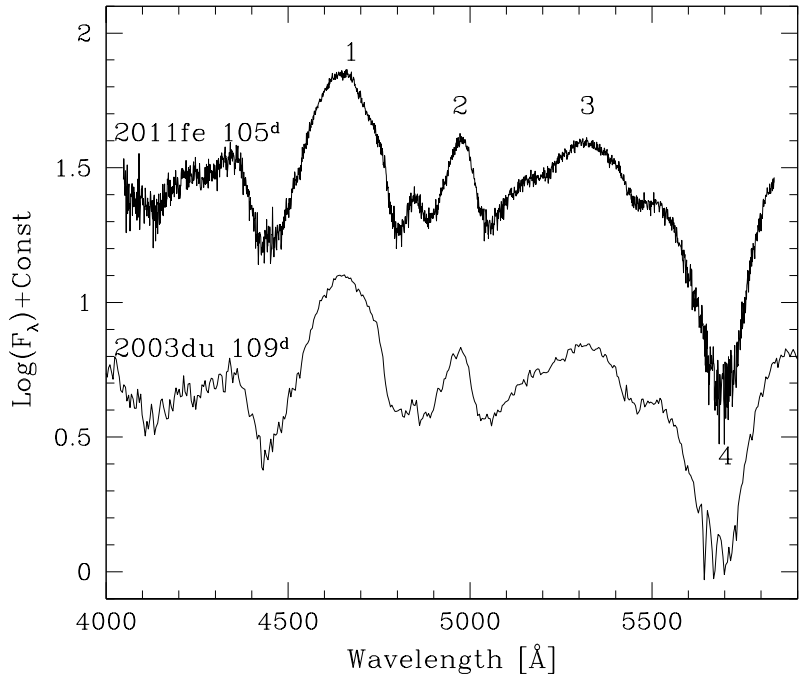}}
\caption{Spectra of SN 2011fe and 2003du.}
\label{f6}
\vspace*{-5.mm}
\end{figure}

The spectrum of SN 2011fe obtained by N600 telescope on December 24.13 UT
(JD 2455919.63, 105 days after the $B$-band maximum) is shown
in Fig.\,\ref{f6}.
The spectrum of SN 2003du obtained 109 days after its
maximum by Stanishev {\it et al.} (2007) is displayed for comparison.
The observed wavelengths are plotted for the
spectrum of SN 2011fe, while the spectrum of SN 2003du was
deredshifted using $z=0.006384$.

\vspace*{-1.cm}

The two spectra are nearly identical, confirming
the physical resemblance of both SNe. This phase in the evolution of
type Ia SNe is considered as transitional between photospheric and
nebular stages. The identification of prominent emission peaks was
proposed by Bowers {\it et al.} (1997): features labeled 1,2 and 3 are blends
of forbidden lines of [Fe III] and [Fe II].
We measured the wavelength centroid of the peaks in our spectrum:
1 -- 4646\AA, 2 -- 4971\AA, 3 -- 5309\AA.
The absorption feature 4 is attributed to Na I by Branch {\it et al.} (2008).
The wavelength of this line in our spectrum is 5697\AA\, which
corresponds to the expansion velocity of about 10100\,km\,s$^{-1}$.

\section{Conclusions}

We present the light and colour curves of SN 2011fe during
652 days after its discovery, as well as one spectrum taken
105 days after its $B$-band maximum. The photometry was 
carried out at five sites,
with 8 telescopes, equipped with different CCD cameras and filter sets.
We used linear colour-terms to transform photometry to the
standard Johnson-Cousins system. It is well known that these colour
corrections do not work well for SNe, because the SN spectral energy
distribution is different from that of normal stars, and the so-called
"$S$-correction" method is preferrable (see, e.g., Stanishev {\it et al.},
2007). Another sources of errors were the use of different comparison
stars at different telescopes and possible field errors. Nevertheless,
the results were found to be generally consistent when comparing
data from different telescopes from our set and comparing our data
with that of other authors.

The light and colour curves for SN 2011fe show that it belongs to the
"normal" subset of type Ia SNe and is almost unreddened.
The decline rate parameter $\Delta m_{15}(B)=1.10$ is close
to the mean value for SNe Ia (see, e.g., Wang {\it et al.}, 2008).
The comparison of light, colour curves and spectrum show that
SN 2011fe is nearly identical in all observed parameters to the
well-studied "normal" SNIa 2003du during the first $\sim$200 days
of evolution.

We found out that the rate
of brightness decline in the {\sl BVR} bands is higher than
average for SNe Ia in the interval of phases 180--430 days.
Afterwards, the decline for $V$, $R$ and "quasi-bolometric" light curves
slows down. This may be caused by the emergence of the light echo,
but some models of late-time luminosity evolution also predict
slowing--down of the decline at that phase (see, e.g., Milne {\it et al.}, 2001).

\acknowledgements
The work of DT and NP was partly supported by the RFBR
grant 13-02-92119. SSh and NK acknowledge support by
the grant of the President of RF No. NSh-2374.2012.2.
In 2012 NK was supported by the National Scholarship
Program (SAIA) of the Slovak Republic.
IV was supported by the RFBR grant 11-02-01213a
and by the National Scholarship
Program (SAIA) of the Slovak Republic.
This work has been supported by the Slovak Academy of Sciences
VEGA Grant No. 2/0002/13 and RFBR grant 11-02-00258a.

We are grateful to D. Chochol for constructive suggestions, which helped
to improve the presentation.

\end{document}